\documentclass{book}    
\usepackage{piers}  
\usepackage{graphicx}
\graphicspath{{./images/}}

\begin{document}

\title{Ray Tracing in an Arbitrary Cloak in Two Dimensions}
\maketitle

\author      {F. M. Lastname}
\affiliation {University}
\address     {}
\city        {Boston}
\postalcode  {}
\country     {USA}
\phone       {345566}    
\fax         {233445}    
\email       {email@email.com}  
\misc        { }  
\nomakeauthor

\author      {F. M. Lastname}
\affiliation {University}
\address     {}
\city        {Boston}
\postalcode  {}
\country     {USA}
\phone       {345566}    
\fax         {233445}    
\email       {email@email.com}  
\misc        { }  
\nomakeauthor

\begin{authors}

{\bf H. H. Sidhwa}$^{1}$, {\bf R. P. Aiyar}$^{2}$, {\bf and S. V. Kulkarni}$^{1}$\\
\medskip
$^{1}$Department of Electrical Engineering, Indian Institute of Technology Bombay, India\\
$^{2}$ CRNTS, Indian Institute of Technology Bombay, India 
\end{authors}

\begin{paper}

\begin{piersabstract}
Electromagnetic wave behaviour in an anisotropic medium with a two dimensional arbitrary
geometry is studied. The aim is to trace the path of a ray in such a complex medium for the purpose
of achieving cloaking (invisibility). A coordinate transformation is carried out for
the formulation of an annular region at the centre of the structure whose topology is a scaled
geometry of outer boundary of the structure.

\end{piersabstract}

\psection{Introduction}
The idea of invisibility has intrigued and fascinated mankind for thousands of years. The motive  is to bend the path of a ray around the body of interest
so as to give an impression that the ray is travelling in a straight line uninterrupted by any obstacle.
The idea of using coordinate transformation for altering the material characteristics which  would in turn cause a change in the formulation 
of Maxwell equations was first espoused by Pendry et al. \cite{pendry_first} and Leonhardt \cite{leo} in 2006. A process of ray tracing in transformed media for spherical and cylindrical cloaks
using Cartesian  tensors was carried out by Pendry et al. \cite{pendry}. A generalised method for designing arbitrarily shaped cloaks using transformation of coordinates approach has been discussed by Li and Li \cite{chen}.
A full wave simulation using a commercial software is carried out to verify the method.\\
\\
In this paper, an attempt has been made to carry out a ray tracing process for an arbitrary  two dimensional cloak.  The region to be cloaked is represented as an arbitrary closed
curve in a two dimensional space. For the purpose of cloaking the region inside the domain is to be mapped to an annular
region similar to the approach given by Li and Li \cite{chen}. A coordinate transformation is carried out for
the formulation of an annular region at the centre of the structure with outer and inner boundaries having same shape.
The formulation for the Hamiltonian is done as reported in \cite{pendry}. Expressions for permittivity and permeability in the anisotropic medium in
the transformed medium are calculated. For verification of the method, it is assumed that the arbitrary
geometry is elliptical in nature. The Hamiltonian is formulated  by
including terms dependent on cylindrical polar coordinate, $\theta$.

\psection{Analysis of two Dimensional Arbitrary Cloak}
A point in space can be expressed by $(r, \theta, z )$ in cylindrical coordinates \cite{chen}.

\begin{align*}
 x&=r\; \cos\theta  \quad  y=r\; \sin\theta 
 \end{align*}
 In a 2 dimensional geometry, let the shape of the outer contour be arbitrary in nature. Let $r=R(\theta)$ define the 
 outer contour. We can define a normalised parameter as:
 \begin{align*}
  \rho=\frac{r}{R(\theta)}=\frac{\sqrt{x^2+y^2}}{R(\theta)}
 \end{align*}
For a given path, $\rho$ remains constant
while the wave traverses path of  $360\,^{\circ}\mathrm{}$.

The coordinates can now be expressed as
\begin{align}
 x&=\rho R(\theta)\; \cos\theta \quad y=\rho R(\theta)\; \sin\theta 
 \end{align}
 For the process of cloaking, we define a transformation which compresses any point lying in the region $0\leq \rho \leq 1 $ to $\tau\leq \rho' \leq 1 $.
 The new region formed is an annular surface with coordinate system $(\rho', \theta', z')$.
 The relationship between the original and transformed coordinate systems can be expressed as:
 \begin{align*}
 0<\rho<1 &\Rightarrow \tau<\rho'<1 \\
\therefore  \rho'&=\tau + (1-\tau)\rho \quad \theta'=\theta \quad z'=z 
 \end{align*}
 Outside the cloaked region lies free space for which corresponding characteristics can be used.
 The corresponding transformed Cartesian coordinates can now be expressed as \cite{chen}:
 \begin{align}
 x'&=r'\; \cos\theta'=\rho'\; R(\theta')\cos\theta\\
 y'&=r'\; \sin\theta'=\rho'\; R(\theta')\sin\theta\\
 x'&=[\tau R(\tan^{-1}\frac{y}{x})+(1-\tau)\sqrt{x^2+y^2}]\frac{x}{\sqrt{x^2+y^2}} \\
 y'&=[\tau R(\tan^{-1}\frac{y}{x})+(1-\tau)\sqrt{x^2+y^2}]\frac{y}{\sqrt{x^2+y^2}}\\
 z'&=z
 \end{align}
 Relative permittivity and permeability tensors can also be transformed as \cite{post}:
 \begin{align}
\epsilon^{i^{'}j{'}}&=\arrowvert\Lambda^{i^{'}}_i\arrowvert^{-1} \Lambda^i{'}_i\Lambda^j{'}_j\epsilon^{ij} \\
\mu^{i^{'}j{'}}&=\arrowvert\Lambda^{i^{'}}_i\arrowvert^{-1} \Lambda^i{'}_i\Lambda^j{'}_j\mu^{ij} \\
\epsilon^{x^{'}x{'}}&=\arrowvert\Lambda^{x^{'}}_i\arrowvert^{-1} \Lambda^x{'}_i\Lambda^x{'}_j\epsilon^{ij}=\arrowvert\Lambda^{x^{'}}_x\arrowvert^{-1} \Lambda^x{'}_x\Lambda^x{'}_x\epsilon^{xx}+
\arrowvert\Lambda^{x^{'}}_y\arrowvert^{-1} \Lambda^x{'}_y\Lambda^x{'}_y\epsilon^{yy}\\
\epsilon^{x^{'}y{'}}&=\arrowvert\Lambda^{x^{'}}_i\arrowvert^{-1} \Lambda^x{'}_i\Lambda^y{'}_j\epsilon^{ij}=\arrowvert\Lambda^{x^{'}}_x\arrowvert^{-1} \Lambda^x{'}_x\Lambda^y{'}_x\epsilon^{xx}+
\arrowvert\Lambda^{x^{'}}_y\arrowvert^{-1} \Lambda^x{'}_y\Lambda^y{'}_y\epsilon^{yy}\\
\epsilon^{y^{'}y{'}}&=\arrowvert\Lambda^{y^{'}}_i\arrowvert^{-1} \Lambda^y{'}_i\Lambda^y{'}_j\epsilon^{ij}=\arrowvert\Lambda^{y^{'}}_x\arrowvert^{-1} \Lambda^y{'}_x\Lambda^y{'}_x\epsilon^{xx}+
\arrowvert\Lambda^{y^{'}}_y\arrowvert^{-1} \Lambda^y{'}_y\Lambda^y{'}_y\epsilon^{yy}\\
\mu^{x'x'}&=\epsilon^{x'x'}\\
\mu^{x'y'}&=\epsilon^{x'y'}\\
\mu^{y'y'}&=\epsilon^{y'y'}
\end{align}
The final expressions for the permittivity in the transformed coordinate system can be seen in \cite{chen}.

\psection{Calculation of Hamiltonian}
Since there is no loss of energy while the wave propagates through the medium,  
the Hamiltonian (given by the following expression), can be found in order to trace the path of the wave in the cloaked medium \cite{pendry}.
\begin{align}
 H&=\frac{1}{2}(1-\tau)(\textbf{k}\textbf{n}\textbf{k}-|\textbf{n}|) \label{eq:hamilt}\\
 \arrowvert \mathbf{n}\arrowvert&= \arrowvert \boldsymbol{\epsilon}\arrowvert=\frac{1}{(1-\tau)^2}\frac{r'-\tau R}{r'}\\
 H&=k_x^2 \epsilon^{x'x'}+2 k_x k_y\epsilon^{x'y'}+k_y^2 \epsilon^{y'y'}+k_z^2 \epsilon^{z'z'}-\arrowvert \epsilon \arrowvert \label{eq:hamiltonian}
\end{align}
where $n$ is the refractive index of the medium and $k$ is the propagation vector.
The path can be parametrised as \cite{orlov}:
\begin{align*}
\frac{d\mathbf{x}}{d\varsigma}=\frac{\partial H}{\partial \mathbf{k}} \\
\frac{d\mathbf{k}}{d\varsigma}=-\frac{\partial H}{\partial \mathbf{x}}
\end{align*}
where $\varsigma$ is the parameterising varaible and $x$ is the position vector.\\
\\
The  Hamiltonian can be solved as per \cite{orlov} as:
\begin{align}
\frac{\partial H}{\partial k_x}&=2 k_x\epsilon^{x'x'}+2 k_y\epsilon^{y'y'}\\
\frac{\partial H}{\partial k_y}&=2 k_x\epsilon^{x'y'}+2 k_y\epsilon^{y'y'} \\
\frac{\partial H}{\partial k_z}&=2 k_z\epsilon^{z'z'}\\
\frac{\partial H}{\partial x}&=k_x^2 \frac{\partial \epsilon^{x'x'}}{\partial x}+2 k_x k_y \frac{\partial \epsilon^{x'y'}}{\partial x}+
k_y^2 \frac{\partial \epsilon^{y'y'}}{\partial x}+k_z^2 \frac{\partial \epsilon^{z'z'}}{\partial x}-
\frac{\partial\arrowvert \boldsymbol{\epsilon}\arrowvert}{\partial x}\\
A&=[(r'-\tau R)^2+\tau^2(\frac{\partial R}{\partial \theta})^2]\cos^2\theta-2\tau r'\frac{\partial R}{\partial \theta}\sin\theta \cos\theta +r'^2 \sin^2\theta \\ 
\epsilon^{x^{'}x{'}}&=\frac{1}{r'(r'-\tau R)}A\label{eq:epxx}\\
\frac{\partial \epsilon^{x'x'}}{\partial x}&=\sin\theta \frac{1-\tau}{r'(r'-\tau R)^2}[\frac{\partial R}{\partial \theta} \frac{\rho'-\tau}{r'-\tau R}2A-\frac{\partial A}{\partial \theta}]
\end{align} 

The expressions for ${\partial \epsilon^{x'x'}}/{\partial y}$ , ${\partial \epsilon^{y'y'}}/{\partial x}$ , ${\partial \epsilon^{y'y'}}/{\partial y}$ , ${\partial \epsilon^{x'y'}}/{\partial x}$ and ${\partial \epsilon^{x'y'}}/{\partial y}$ can be calculated in a similar way for the calculation of 
${\partial H}/{\partial y}$.\\ \\
For the verification of the above algorithm, the outer contour is considered to have the shape of an ellipse with major axis $a$ and minor axis $b$.
\begin{align}
 R&=\frac{ab}{\sqrt{b^2 \cos^2 \theta+a^2 \sin^2 \theta}}\\
 \frac{\partial R}{\partial \theta}&=\frac{\sin\theta cos\theta (b^2-a^2)ab}{(b^2 \cos^2 \theta+a^2 \sin^2 \theta)^{3/2}}\\
\frac{\partial^2 R}{\partial \theta^2}&=(b^2-a^2)ab(\frac{\cos 2\theta}{(b^2 \cos^2 \theta+a^2 \sin^2 \theta)^{3/2}}+\frac{3}{4}\frac{(b^2-a^2)\sin^2 2\theta}{(b^2 \cos^2 \theta+a^2 \sin^2 \theta)^{5/2}})
 \end{align}

On substituting for R and its derivatives in Eq. (\ref{eq:hamiltonian}) for the Hamiltonian, and solving for position vector $x$ and propagation vector $k$, 
the path of the ray inside the cloaked medium is traced.

\psection{Results for an elliptical cloak}

\begin{figure}[h!]
\begin{center}
\includegraphics[width=0.8\textwidth]{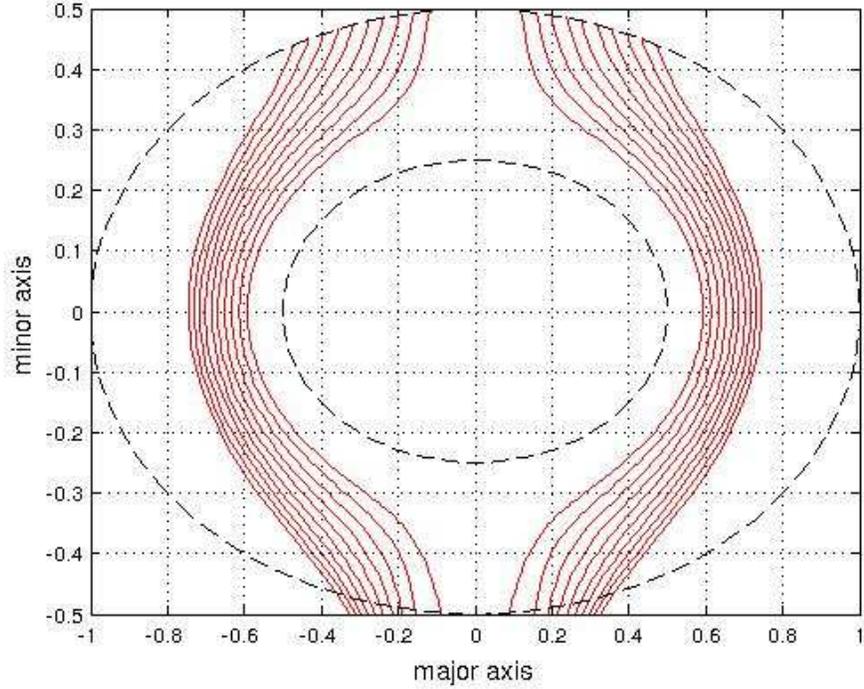}
 \caption{Plot of elliptical cloak  in an X-Y plane  for $\frac{a}{b}=2, \: \: \tau=0.5  $}
 \label{fig:2Dcloak}
 \end{center}
     \end{figure}

 \begin{itemize}
  \item The elliptical cloak in fig.\ref{fig:2Dcloak} shows that a medium with properties defined by Eq. (\ref{eq:epxx}) would exhibit cloaking for
electromagnetic waves in geometric limit (the wavelength $\lambda\ll a$ and $\lambda\ll b$ in the case of ellipse).
\item Experimentally, these structures cannot be realized using naturally occurring materials, but on using metamaterials which have properties like Eq. (\ref{eq:epxx}),
i.e. variable $\epsilon_r,\; 0<\epsilon_r<1$  etc, one can  realise such characteristics using split ring resonators or complementary split ring resonators.
\end{itemize}

%
%
%

\psection{Conclusion}
A generalised method for ray tracing of an arbitrarily shaped cloak in two dimensions by finding the Hamiltonian  is illustrated.  
In the existing literature, formulations have been given for specific shaped cloaks and verified by ray tracing. Recently, algorithms
have been proposed for arbitrarily shaped cloaks, but these algorithms have not been tested through ray tracing method.
In this paper, one such algorithm has been used for realising an arbitrarily shaped cloak. The method has been verified through ray tracing.
Any arbitrary two dimensional closed curve can be parametrised by specifying $\rho$\ as a function of $\theta$. The function $\rho(\theta)$ 
must be a single valued function of $\theta$ for the described algorithm to work efficiently. Thus any curve satisfying the above characteristics can be cloaked
and verified by the process of ray tracing.

%
 \ack
The authors would like to thank Prof. R. K. Shevgaonkar for his valuable suggestions throughout this work. The help given by Mr.  Chinmay Rajhans 
towards the mathematical formulation is also appreciated.

\end{paper}

\end{document}